\documentclass{aa}
\usepackage{graphicx}
\begin{document}
\title{First results of a cryogenic optical photon counting imaging
spectrometer using a DROID array}

\author{R.A. Hijmering\inst{1}
\and P. Verhoeve\inst{1} \and D.D.E. Martin\inst{1} \and R.
Venn\inst{2} \and A. van Dordrecht\inst{1} \and  P.J. Groot\inst{3}}

   \subtitle{}

\institute{Advanced Studies and Technology Preparation Division,
Directorate of Science and Robotic Exploration of the European Space
Agency, P.O. Box 299, 2200 AG Noordwijk, The Netherlands\\
              \email{rhijmeri@rssd.esa.int}
\and Cambridge MicroFab Ltd., Broadway, Bourn, Cambridgeshire CB3
7TA, UK \and Department of Astrophysics, Radboud University
Nijmegen, P.O. Box 9010, 6500 GL Nijmegen, The Netherlands}

\date{Received August 20, 2009 /
Accepted <date>}

  \abstract
{In this paper we present the first system test in which we
demonstrate the concept of using an array of Distributed Read Out
Imaging Devices (DROIDs) for optical photon detection.}
{After the successful S-Cam 3 detector the next step in the
development of a cryogenic optical photon counting imaging
spectrometer under the S-Cam project is to increase the field of
view using DROIDs. With this modification the field of view of the
camera has been increased by a factor of 5 in area, while keeping
the number of readout channels the same.}
{The test has been performed using the flexible S-Cam 3 system and
exchanging the 10x12 Superconducting Tunnel Junction array for a
3x20 DROID array. The extra data reduction needed with DROIDs is
performed offline.}
{We show that, although the responsivity (number of tunnelled
quasiparticles per unit of absorbed photon energy, e-/eV) of the
current array is too low for direct astronomical applications, the
imaging quality is already good enough for pattern detection, and
will improve further with increasing responsivity.}
{The obtained knowledge can be used to optimise the system for the
use of DROIDs.}

   \keywords{detectors:photometers: spectographs: spectrometer}

\authorrunning{R.A. Hijmering et al.}
\titlerunning{First results of a cryogenic optical photon counting$\cdots$}
   \maketitle

\section{Introduction}\label{intro}
With the S-Cam project the Advanced Studies \& Technology
Preparation Division of the European Space Agency is developing a
series of prototype cryogenic detectors to be used as optical photon
counting imaging spectrometers for ground-based astronomy. S-Cam
uses Superconducting Tunnel Junctions (STJs) (\cite{Friedrich:2006};
\cite{Prober:2006}; \cite{Peacock:1996}) as its detector technology.
The merit of this and other cryogenic detectors (\cite{Romani:1999})
is that they combine single photon detection with sub-microsecond
time resolution and intrinsic wavelength resolution, imaging and
good detection efficiency in a single device.

STJs consist of 2 superconducting layers separated by a thin
insulating layer acting as a tunnel barrier. With the absorption of
a photon in the superconducting layer a large quantity (several
thousands) of Cooper pairs are broken into quasiparticles which can
tunnel across the barrier and, under the influence of an applied
bias voltage, produce a measurable current pulse. The number of
created quasiparticles is given by:
$N(E_{0})=\frac{E_{0}}{\varepsilon}$, with $N(E_{0})$ the number of
created quasiparticles, $E_{0}$ the energy of the absorbed photon
and $\varepsilon=1.75\Delta_{g}$ the mean energy needed to create a
quasiparticle (\cite{Kurakado:1981}) with $\Delta_{g}$ the energy
gap of the superconducting material. As shown the number of created
quasiparticles, and hence the amplitude of corresponding tunnel
current, is proportional to the energy of the absorbed photon, thus
providing the detector with its spectrographic capabilities. The
theoretical limit for the intrinsic energy resolution is given by:
$\Delta E= 2.355\sqrt{\varepsilon E_{0} (F+G)}$, where $F$ is the
Fano factor (\cite{Fano:1947}), equal to $F=0.2$
(\cite{Kurakado:1981}; \cite{Rando:1991}), and $G=1+\frac{1}{<n>}$
(\cite{Mears:1993}) accounts for the statistical variations in the
tunnel process, with $<n>$ the average number of tunnels of a single
quasiparticle. The energy gap, $\Delta_{g}$, of the superconducting
material is proportional to its critical temperature, $T_{c}$), the
temperature at which the phase changes from superconducting to
normal metal. For a BCS type superconductor (usually an elemental
superconducting material which follows the theory developed by
\cite{Bardeen:1957}), $\Delta_{g}=1.764k_{b}T_{c}$. A lower energy
gap of the superconducting material will therefore increase the
number of created quasiparticles and provide better spectrographic
capabilities, but it also puts increasing constraints on the
operating temperature ($T_{op}$). This needs to be well below the
critical temperature of the superconducting layer
($T_{op}\approx0.1T_{c}$) in order to sufficiently reduce the
thermally excited quasiparticle population. For a more  extended
overview of the STJ technology the reader is referred to
\cite{Peacock:1996}.

Each STJ needs to be read out using a dedicated electronics chain
which limits the maximum number of pixels that can be read out in a
practical application (\cite{Martin:2006}). To overcome this
limitation the Distributed Read Out Imaging Device (DROID)
(\cite{Kraus:1989}) is being developed. A DROID consists of a
superconducting absorber strip with STJs on either end, see fig.
\ref{fig:Figure 1}. The photon is absorbed in the absorber and the
created quasiparticles diffuse towards the STJs where they tunnel.
The sum of the tunnel signals of both STJs is a measure for the
energy of the absorbed photon while the ratio is a measure for the
absorption position. Depending on the position resolution of the
DROID it can replace a number of single STJs and reduce the number
of read out channels for a given sensitive area
(\cite{Hijmering:2008}).

Within the S-Cam project three prototype cameras have already
successfully been used on telescopes such as the William Herschel
Telescope (WHT) on La Palma and the Optical Ground Station (OGS) on
Tenerife (\cite{Martin:2004}). S-Cam 1 (\cite{Verhoeve:2002}) and 2
(\cite{Rando:2000}) were based on a $6\times6$ pixel array
($25\times25 \mu m^{2}$ pixels) with a wavelength resolving power of
6. S-Cam 3 (\cite{Martin:2007}; \cite{Martin:2006}) was based on a
$10\times12$ pixel array ($35\times35 \mu m^{2}$ pixels), increasing
the field of view on the WHT from $4\arcsec\times4\arcsec$ to
$10\arcsec\times12\arcsec$. Also the covered wavelength range,
operating temperature and resolving power ($\sim14@500nm$) has been
enhanced with S-Cam 3. The applicability of this type of detector
has been proven in different observation campaigns in which several
types of astronomical objects have been observed. The high time
resolution spectrally resolved S-Cam data has provided strong
constraints on the geometry of eclipsing binaries
(\cite{Perryman:2001}; \cite{de Bruijne1:2002}; \cite{Martin:2003}).
Precise timing of the Crab pulsar light curve has shown that the
optical pulses lead the radio pulses by $273\pm65\mu s$
(\cite{Perryman:1999}; \cite{Oosterbroek:2006}). The spectral
information provided by the STJs has enabled the direct
determination of quasar redshifts (\cite{de Bruijne2:2002}) and
stellar temperatures (\cite{Reynolds:2003}). The next step is to
 increase the field of view further with the use of DROIDs. Here we
present the results of the first system test using a $3\times20$
DROID array as a detector.

\section{Operation of the DROID array}\label{operation}

\begin{figure}
\centerline{
\includegraphics[width=8.5cm,angle=-0]{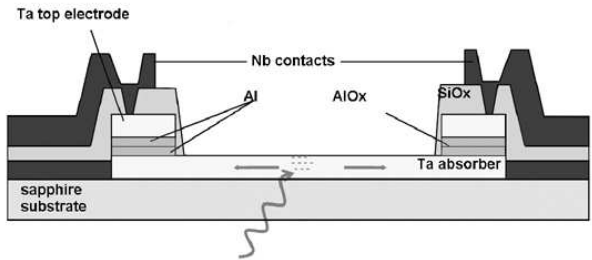}}
\caption[]{Schematic representation of the DROID geometry used in
the DROID array.} \label{fig:Figure 1}
\end{figure}

\begin{figure}
\centerline{
\includegraphics[width=8.5cm,angle=-0]{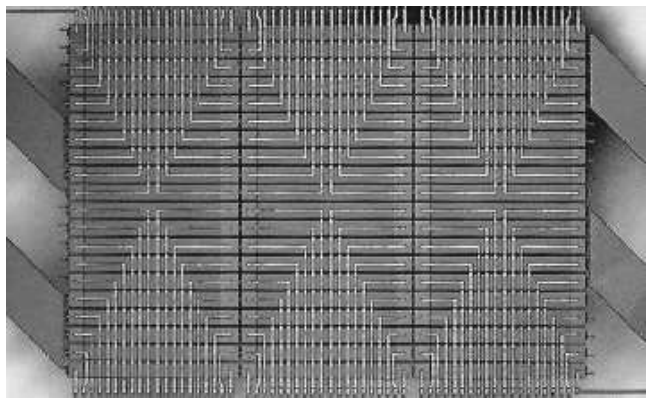}}
\caption[]{The $3\times20$ format DROID array fabricated by MicroFab
Ltd. The leads for the read out of the STJs are visible running over
the absorbers. The array is divided into 4 electrically isolated
groups of $3\times5$ DROIDs of which the common ground leads can be
identified as the two wide strips on either side of the array.}
\label{fig:Figure2}
\end{figure}

The DROID array, shown in fig. \ref{fig:Figure2}, is fabricated by
MicroFab Ltd. and is based on pure tantalum absorbers
($\Delta_{Ta}=700\mu eV$) with proximized Ta/Al STJs on the side
($\Delta_{STJ}=500\mu eV$). The lay-up of the STJs is
Ta/Al/AlOx/Al/Ta with thicknesses of 100/30/1/30/100$nm$. The
100$nm$ thick tantalum absorber of the DROID and the tantalum layer
of the base electrode of the STJ (see fig. \ref{fig:Figure 1}) are
made of a single layer of tantalum. The presence of the aluminium
layer in the STJ reduces the energy gap due to the proximity effect
(\cite{Booth:1987}) and provides confinement of quasiparticles
inside the STJ, which enhances the performance. The confinement of
quasiparticles using this method is not always 100\% effective and
quasiparticles which reside at higher energies,
$\varepsilon_{qp}>\Delta_{Ta}$, are able to escape the STJ into the
absorber. The DROIDs are $33.5\times360\mu m^{2}$ in size, including
the $33.5\times33.5\mu m^{2}$ STJs. The DROIDs are separated by
$4\mu m$ wide gaps to accommodate the interconnections between the
base electrodes of the STJs, which share a common return wire. These
interconnections are made of higher gap material (Nb,
$\Delta_{Nb}=1550\mu eV$) which prevents diffusion of quasiparticle
across the interconnections, and thereby cross talk between DROIDs.
The leads to the top electrodes of the STJs are routed over the
front side of the DROIDs outwards. In order to electrically isolate
the leads to the top electrodes from the rest of the DROID structure
the complete array has been covered with $SiO_{x}$. The array is
divided into 4 electrically isolated groups of $3\times5$ DROIDs,
each with a single common return lead. The devices are made on a
transparent sapphire substrate which allows for backside
illumination, through the sapphire. In this way the wiring routed
over the absorber at the front side does not block any photons.

The individual DROIDs on the array have been characterised in a
$^{3}He$ sorption refrigerator in which two DROIDs can be read out
at a time. In this cryostat the devices are effectively shielded
from IR radiation using a closed shield surrounding the sample space
and the chip can only be illuminated via an optical fibre. The
devices are biased using a small voltage bias and the electronics
used to read out a single DROID at a time consist of two charge
sensitive preamplifiers, with a RC time of $470\mu s$, each followed
by a shaping stage. The two channels are linked in the sense that
coincident events, defined as events in either STJ, resulting from a
photon absorption in the DROID which occur within $30\mu s$ (the
time window is defined by the electronics and the time of arrival is
defined as the time the signal passes through a threshold), can be
identified and selected, while uncorrelated events are rejected.
This efficiently reduces the noise-induced events as well. The
resulting data for each event consists of the pulse height values of
the two channels and the relative time of detection.

\begin{figure}
\centerline{
\includegraphics[width=8.5cm,angle=-0]{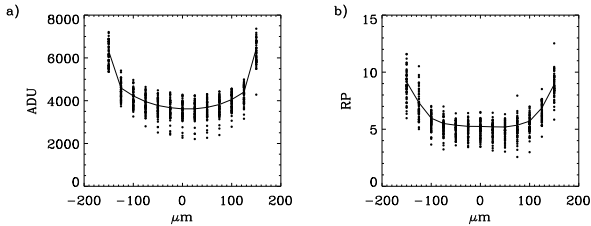}}
\caption[]{a) The relative responsivity for each of the DROIDs in
the array (black dots) and the average responsivity (solid line)
versus position along the absorber. b) The wavelength resolving
power at $400nm$ of the individual DROIDs in the same
representation. The outer group on either side represents the STJs.}
\label{fig:Figure 3}
\end{figure}

The characterisation of the array showed that two DROIDs in the
array were erroneously interconnected and one DROID showed increased
subgap current levels. Fig. \ref{fig:Figure 3}a and b shows the
relative pulse height (sum of the two STJ pulse heights) and
wavelength resolving power $\frac{\lambda}{\Delta \lambda}$,
respectively, of the DROIDs in the array as a function of position
along the absorber (derived from the ratio of the two STJ pulse
heights) under illumination with $\lambda=400nm$ photons. The
relative responsivity and resolving power of the DROIDs are
determined by fitting a Gaussian to the resulting single peak in
pulse height histogram. The results show that the responsivity is
rather low, roughly a order of magnitude lower compared to
previously tested DROIDs. The responsivity across the array is
rather non-uniform, with a standard deviation of 24\%. This problem
is related to the variable quality of the Nb interconnections
between the base electrodes and solutions are currently under
investigation. For practical use the absorber is divided into
sections or virtual pixels. For S-Cam the size of a virtual pixel
will be $33\mu m$, equal to the width of the absorber, and
corresponding to the $\sim1\arcsec$ seeing on the sky at the William
Herschel Telescope and the Optical Ground Station. The average
wavelength resolving power for the absorber events is $6\pm1
(@\lambda=400nm$), and corresponds to an average position resolution
of $\Delta x \approx35\mu m$ (\cite{Jochum:1993};
\cite{Hijmering:2008}), well matched to the size of a virtual pixel.
The $1\sigma$ variation in wavelength resolving power over the array
is 16\% (see fig. \ref{fig:Figure 3}b) and is directly correlated to
the variations in responsivity.

\section{Full array test set-up}\label{setup}

\begin{figure}
\centerline{
\includegraphics[width=8.5cm,angle=-0]{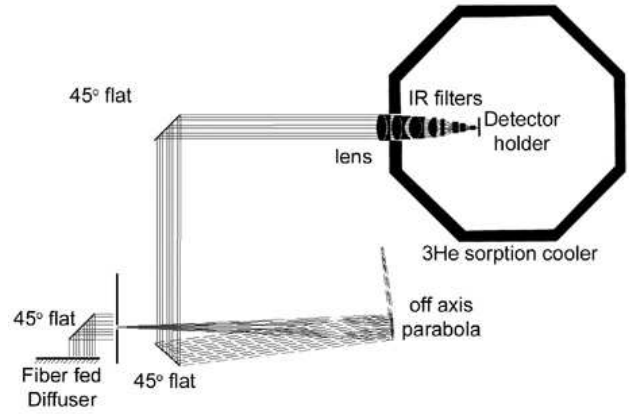}}
\caption[]{Schematic representation of the S-Cam 3 optical set-up.}
\label{fig:Figure 4}
\end{figure}

A full array test has been performed with the S-Cam 3 system
(\cite{Martin:2006}; \cite{Martin:2007}) in which the complete array
can be read out simultaneously and the array can be illuminated from
the outside through a window. The optical chain, fig.
\ref{fig:Figure 4}, consists of an off axis paraboloidal mirror, two
flat $45^{\circ}$ mirrors to fold the beam, a high quality lens
system to focus the beam on the detector and a set of 3 cold IR
filters (inside the cryostat) to reduce the thermal load and IR
background. The optical chain has a demagnification factor of 5.4
and the available wavelength band, limited by the IR filters, is
$345-750nm$. For laboratory tests the focal plane of the off axis
paraboloidal mirror is illuminated, through a diffuser, with
monochromatic light from a Xe lamp and grating monochromator through
a UV-grade optical fibre. A pinhole can be moved into the focal
plane to project spots of various size and shapes onto the chip. The
chip is back-illuminated through the sapphire substrate to avoid
obscuration by the readout leads and to exploit the infrared
absorption properties of sapphire. The cryostat contains a liquid
helium bath and uses a double stage closed cycle $^{3}He/^{4}He$
evaporation cooler with a base temperature of $290mK$ and a hold
time of $\sim28$ hours. The readout, which is similar to the one
used with the measurements on one individual DROID, is performed
using 120 charge sensitive preamplifiers grouped into the four
electrically isolated groups, followed by analogue-to-digital
converters and a programmable Finite Impulse Response (FIR) filter
which acts as a shaping stage. The implemented filters produce a
bipolar output pulse for each detected photon, of which both the
positive and negative peak are sampled for offline evaluation and
the pass through zero of the bipolar signal defines the time of
arrival with a known offset. The ratio of positive and negative peak
amplitude carries information on the original pulse shape and can be
used to discriminate photon induced events from other disturbances.
Each event in the STJ is labeled with a $1\mu s$ accurate time stamp
derived from a GPS (Global Positioning System) signal. The collected
data for each detected event consists of the label of the STJ, the
pulse height values for the positive and negative amplitudes and a
time stamp.

Although the data acquisition system is very flexible it is
currently geared towards the readout of an array of 120 single STJs,
which introduces some complications for the read out of DROIDs. The
signal in one of the STJs of a DROID decreases with distance between
the absorption position and the STJ, and in order to detect the
signals from absorptions near the far side STJ the thresholds for
the individual channels need to be set sufficiently low, which
introduces a large amount of noise-induced events. Identification
and selection of coincident events in the two STJs of a DROID cannot
be performed at a hardware level yet and therefore all triggered
events (including noise triggers) have to be recorded. Coincident
event selection is then performed offline using the time stamps of
the individual events. Implementation of the DROID array and
operation of the system proved to be not more difficult than the
operation of the original 120 pixel S-Cam 3 detector array. Similar
to previous experience, the array was more sensitive to the trapping
of magnetic flux due to the larger superconducting area of the chip
and multiple cool down cycles were required before an optimally
functioning array was obtained. During the measurement four DROIDs
where set inactive to allow stable operation: the two interconnected
DROIDs, the DROID with increased subgap current levels and one DROID
which remained flux trapped. Flux trapping can be removed by heating
up the devices to a temperature above $T_{c}$, removing any remnant
field and cool down. If the magnetic field is low enough no flux
will be trapped. For future use the interconnected DROIDs can be
separated and improved magnetic shielding should remove the flux
trapping entirely leaving only a single bad DROID in the array.
Because of the low responsivity of the DROID array only the shortest
wavelengths in the available wavelength band could be used for
illumination. Even so, the signal for some of the DROIDs did not
reach above the detection threshold. For the current demonstration
this results effectively in a non-uniform efficiency, which can be
reduced using a flat field correction. However, because of the low
responsivity the array is not useful for application on a telescope.

\section{Data reduction}\label{datareduction}
\begin{figure}
\centerline{
\includegraphics[width=8.5cm,angle=-0]{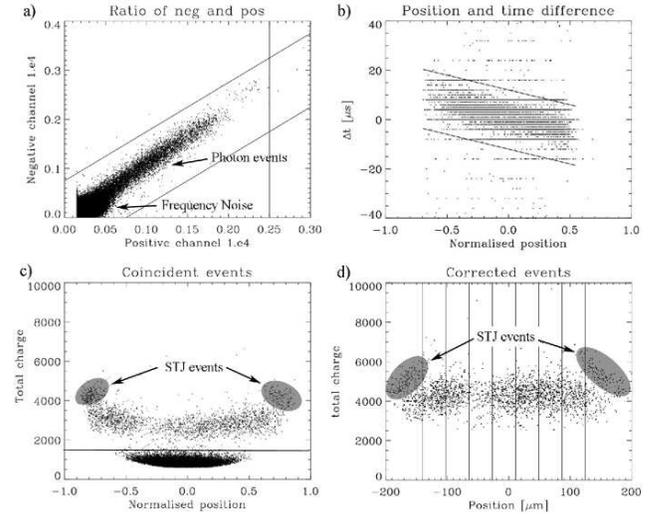}}
\caption[]{Representation of the filtering procedures of the DROID
array data. The data of the individual graphs is taken from
different measurements to aid clarity of the complete process. a)
Filtering on the ratio of the positive and negative peak which
should lie close to unity. The positive channel does not reach down
to zero because the detection threshold is set in this channel. b)
After identifying coincident events filtering on the correlation
between the ratio of charges and difference in time stamp and c)
removing the noise events with low pulse heights using the sum of
the two signals. d) Shows the data after correction using the model
from \cite{Jochum:1993} and divided into different sections. The gap
in the absorber data of graphs c) and d) is caused by a mask in the
focus of the optical system of which the result is shown in fig.
\ref{fig:Figure 8}.} \label{fig:Figure 5}
\end{figure}

Despite the non-uniformity, the data of the DROIDs in the array
shows fairly similar patterns and the off-line data reduction can
easily be automated. The individual event data are initially
filtered on the ratio of the positive and negative peak amplitudes
(fig. \ref{fig:Figure 5}a) which should be close to unity for true
photon absorptions (\cite{Martin:2007}). Coincident events are
defined as events in the two STJs belonging to the same DROID which
occur within $40\mu s$, this time window is set manually in the
offline  data reduction and is optimized for the obtained data. In
this step $\sim95\%$ of the events, mainly noise-induced events, are
filtered out making it the most important filtering step. Even with
current computational power it takes several hours to complete this
filtering on a file of several minutes' acquisition ($\sim35$
million events). The time difference between the two signals depend
on the distance the quasiparticles have to travel towards the STJs.
Because the ratio of the charges is a measure of the absorbtion
position, $\frac{Q_{r}-Q_{l}}{Q_{r}+Q_{l}}$ (with $Q_{l/r}$ being
the pulse height value of the left or right STJ), there is a
correlation between the ratio of charges and time difference. This
correlation can be used as an extra filtering condition (fig.
\ref{fig:Figure 5}b). This step is the second most important step
with an additional $\sim70\%$ rejection efficiency. Finally the
noise events at low pulse height are rejected by setting a lower
threshold on the sum of the two signals $Q_{r}+Q_{l}$, see fig.
\ref{fig:Figure 5}c.

The STJ events can be discriminated from the absorber events by
their spatial and spectral separation (fig. \ref{fig:Figure 5}c).
This is possible because the array is illuminated with monochromatic
light. Part of the STJ events overlap with the outer absorber events
in position and in case a broad band spectrum is used for
illumination the spectral separation disappears. In order to
calculate a correct measure for the photon energy and absorption
position in the absorber the model of \cite{Jochum:1993} is used
(fig. \ref{fig:Figure 5}d). Although this model is not complete in
the description of all the processes involved with photon detection
using DROIDs it provides an adequate and simple reconstruction
method for the absorber events using only 2 fitting parameters
(\cite{Hijmering:2008}). The energy $E_{0}$ and position of
absorption $x_{0}$ are derived from the measured signal amplitudes
$Q_{r}$ and $Q_{l}$ as shown in equations (\ref{eq:1}) and
(\ref{eq:2}).

\begin{equation}\label{eq:1}
E_{0}=c\sqrt{(Q_{r}^{2}+Q_{l}^{2})(1-\beta^{2})+2Q_{r}Q_{l}[(1+\beta^{2})cosh\alpha+2\beta
sinh\alpha]}\nonumber
\end{equation}
\begin{equation}\label{eq:2}
x_{0}=\frac{L}{2\alpha}ln\left(\frac{Q_{r}e^{\frac{\alpha}{2}}(1+\beta)+Q_{l}e^{-\frac{\alpha}{2}}(1-\beta)}{Q_{r}e^{-\frac{\alpha}{2}}(1-\beta)+Q_{l}e^{\frac{\alpha}{2}}(1+\beta)}\right)
\end{equation}

Here $c$ is the conversion factor between the measured charge and
photon energy which can be obtained by calibration. The values for
the fitting parameters $\alpha$, corresponding to loss in the
absorber, and $\beta$, describing the trapping efficiency in the
STJs, are determined from a least squares fit of the model to the
absorber data, thus without the events in the STJs. The resulting
values for $\alpha$ and $\beta$ for the DROIDs in the array are
averaged to obtain a single value for $\alpha=1.4\pm0.3$ and
$\beta=0.4\pm0.2$ which are then used for the correction. This is
possible due to the homogeneity of the loss in the absorber and
confinement of quasiparticles in the STJ in the DROIDs across the
array. After this reconstruction is applied the absorber events are
divided into 9 sections which are of roughly the same size as the
STJs. The edges of the absorber in the data and the STJ events are
determined by eye and the data points in between are separated in
equidistant section. Each section will represent a virtual pixel in
the final images.

\section{Imaging quality of the DROID array detector}\label{quality}

\begin{figure}
\centerline{
\includegraphics[width=8.5cm,angle=-0]{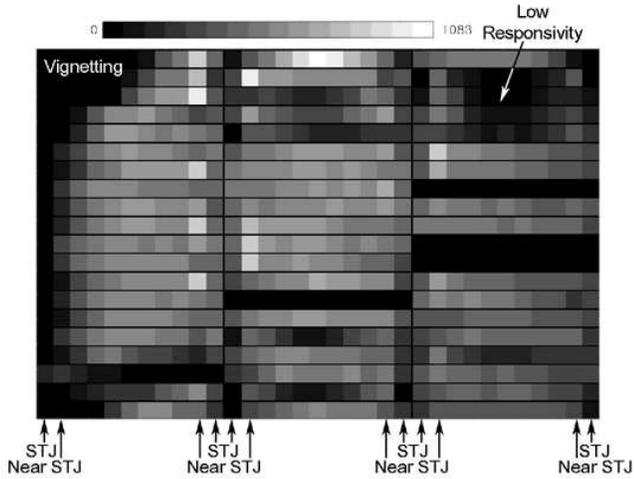}}
\caption[]{Image under full illumination used for the flat field
correction. Each DROID corresponds to 11 virtual pixels in the
horizontal direction. The left hand side shows dark areas due to
vignetting. The DROIDs (2,7), (3,9), (3,10) and (3,13) are switched
off.} \label{fig:Figure 6}
\end{figure}

An image of a full illumination of the array ($\lambda=345nm$, the
lower limit of the available wavelength band) is shown in fig.
\ref{fig:Figure 6} illustrating the non-uniform response of the
array. The left hand side of the image is affected by vignetting
from the lens assembly. This is more prominent compared to all
images shown below due to a shift sideways of the lens assembly due
to a re-alignment. Four DROIDs were switched off and appear as 11
black pixels in a row, one in the second column, line 7, and 3 in
the third column, lines 9, 10 and 13. The remaining non-uniformity
is caused by the low responsivity of the devices in the following
manner. The responsivity of some devices, such as in the upper right
corner, is too low to lift the sum signal from an optical photon
absorbed in the middle of the DROID above the threshold set to
reject the coincident noise triggers (see fig. \ref{fig:Figure 5}c)
and these events are erroneously rejected as noise, producing dark
areas in the image. If more energetic photons were used the signals
of all events would rise above the thresholds and a much more
uniform flat field would be obtained. The same holds for an array
with higher responsivity. As long as the threshold settings as well
as the wavelength of illumination remain unchanged, the above image
can be used for flat field corrections on other images. At the
positions where the flat field shows no events, e.g. due to
vignetting, the correction factor is set to unity and no correction
is applied.

\begin{figure}
\centerline{
\includegraphics[width=8.5cm,angle=-0]{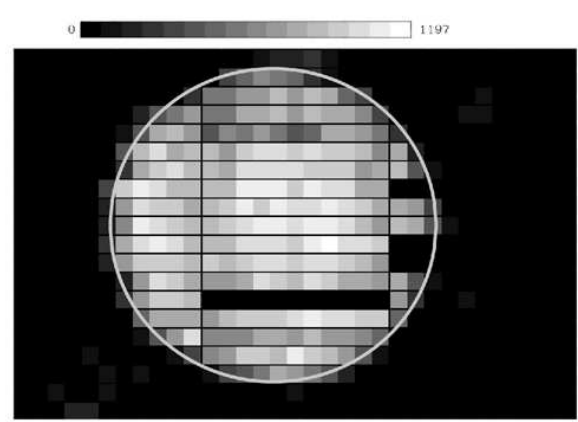}}
\caption[]{Image of an illumination through a $3.4 mm$ aperture in
the focus (after flat field correction with the image of fig.
\ref{fig:Figure 6})} \label{fig:Figure 7}
\end{figure}

The imaging capabilities have been tested by illuminating the array
through a set of masks positioned in the focus of the off-axis
paraboidal mirror. Fig. \ref{fig:Figure 7} shows a reconstructed
image for the case when the array was illuminated through a $3.4mm
\diameter$ (diameter) aperture in the focus of the off axis
paraboloidal mirror, which should be projected as a $630\mu m
\diameter$ image on the detectors. The circle, which represents the
predicted size of the image, overlaps the boundaries of the image
indicating correct scaling and the sharp drop-off of the intensity
at the edges suggests a correctly focussed image and good position
resolution along the DROID.

\begin{figure}
\centerline{
\includegraphics[width=8.5cm,angle=-0]{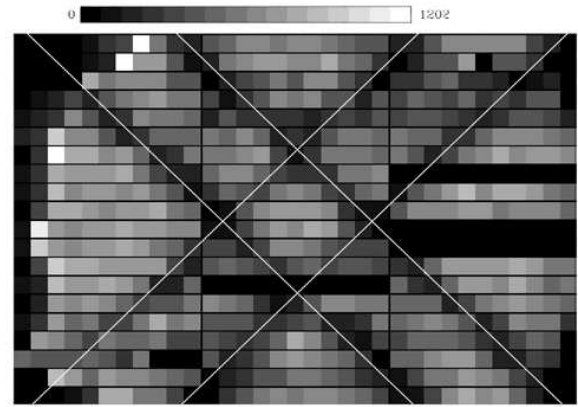}}
\caption[]{Image of an illumination through a mask with of a double
cross pattern with line width of $0.1mm$ and $1.1mm$ separation
(after flat field correction with the image of fig. \ref{fig:Figure
6})} \label{fig:Figure 8}
\end{figure}

Secondly a mask with a double cross structure with line width of
$0.1mm$, corresponding to $16\mu m$ on the detector, was used. The
spacing between the lines is $1.1mm$ in the focus which should
result in a spacing of $204\mu m$ at the detector. Fig.
\ref{fig:Figure 8} shows the resulting intensity plot with the
projected image of the double cross represented by the lines. On the
image the shadows of the double cross can be identified reasonably
well and the lines overlap, showing the correct scaling. S-Cam is
being developed for applications in ground based astronomy.

\begin{figure}
\centerline{
\includegraphics[width=8.5cm,angle=-0]{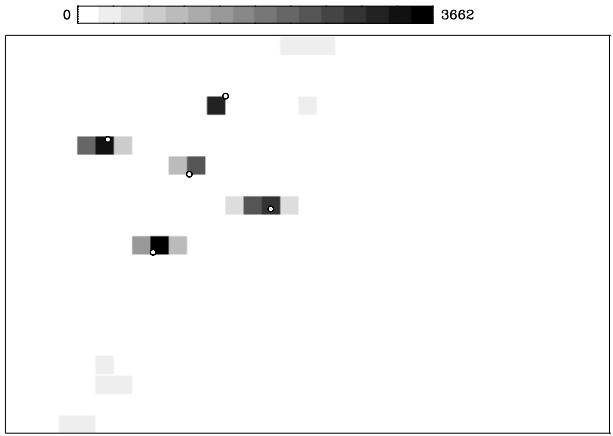}}
\caption[]{Negative image of the 5 pinhole pattern. The five circles
which represent the predicted pattern on the detector.}
\label{fig:Figure 9}
\end{figure}

In order to simulate illumination from the sky a pinhole pattern has
been located in the focus of the off axis parabola. Five $50\mu m
\diameter$ pinholes are located on an $850\mu m$ grid which should
result in five $10\mu m$ spots, close to the limit of the optics
resolution, on a $160\mu m$ grid on the detector. Fig.
\ref{fig:Figure 9} shows the resulting image in negative. The
pinhole pattern is deliberately projected on an area of the array
with good responsivity. The predicted size and positions of the
spots on the array are plotted over the image. The spots do not
perfectly overlap because the position resolution in the vertical
direction is determined by the width of the absorber. The upper and
middle points show less broadening along the DROID length. These
spots are located directly on a STJ where, due to the lower energy
gap and quasiparticle confinement, the energy and position
resolution is improved. The broadening of the other three points is
slightly above a virtual pixel and corresponds with the position
resolving power of $35\mu m$ shown in section \ref{operation} as
derived from the energy resolving power.

\section{Discussion}\label{discussion}
We have successfully demonstrated the feasibility of a DROID array
as an imaging detector in the UV/visible. The S-Cam 3 $10\times12$
individual STJ array could easily be exchanged for the $3\times20$
DROID array without introducing extra operating difficulties into
the system. However, the larger superconducting area of the array
appears to be more sensitive to trapping magnetic flux and better
shielding inside the cryostat is required. The responsivity of the
current array is too low for use on the telescope and only the
shortest wavelengths in the available range of $340-740nm$ could be
used. The responsivity is also non-uniform over the array, such that
for some DROIDs not even the shortest wavelengths could be detected
over the entire absorber. In order to reduce the effect on the
images a flat field correction has been applied. The position
resolution is found to be slightly larger than the absorber width as
shown by illuminations with point sources as shown by fig.
\ref{fig:Figure 9}, and agrees with the position resolution of
$\sim35\mu m$ calculated from the energy resolving power. This
position resolution is just adequate to replace 11 individual STJs
with a single DROID. This corresponds to an array of 660 virtual
pixels using only 120 readout channels which would amount to a field
of view of $20\arcsec\times30\arcsec$ on the William Herschel
Telescope. The imaging capability of the DROID array has been
demonstrated by using a $3.4mm$ aperture and a $1.1mm$ separated
double cross with $0.1mm$ wide lines, fig. \ref{fig:Figure 7} and
fig. \ref{fig:Figure 8}. Both images show a recognisable image of
the introduced object as expected from the good position resolution.
The obvious first improvement on the DROID array for use on the
telescope would be to increase the responsivity, which will
automatically improve the energy and position resolution. This can
be achieved by reducing the loss of quasiparticles inside the
system. Experience has shown that the problem of low and variable
responsivity appears to be related to the quality of the niobium
interconnections between adjacent DROIDs and solutions are under
investigation. In addition, increasing the thickness of the
aluminium trapping layers in the STJ would improve the trapping of
the quasiparticles in the STJs which will improve the responsivity
thereby both energy and position resolution as well. Such
performance has already been demonstrated in measurements on single
DROIDs of identical geometry in \cite{Hijmering:2006} and
\cite{Hijmering:2009}. An aluminium layer thickness of 60nm is
advised, which is a good trade off between the trapping in the STJs
and thermal noise with the current base temperature of just below
300mK. There is also margin for decreasing the base temperature with
more advanced sorption coolers which will allow even thicker Al
layers to be considered. In addition to improvements on the DROID
array some practical improvements can be introduced to the data
acquisition system. Most important is to link the two readout chains
for each DROID making it possible for the electronics to detect
coincident events. If an event in one STJ is followed by an event in
the other STJ within a user-determined time window the event is
passed on as valid, otherwise it is discarded. Secondly, because the
threshold on a single channel must be set low in order to detect the
low signals in case of absorption near the opposite STJ a lot of
coincident noise events will be passed. These can be effectively
filtered out by introducing an extra threshold on the sum of the two
pulse heights, effectively setting an upper wavelength limit for
detection. These two modifications to the system would improve noise
filtering and offline data reduction time by orders of magnitude.
Finally the data could be converted into position and energy data
instead of the charges of the STJs, by using for instance the model
from \cite{Jochum:1993}. However care has to be taken where this is
implemented. In order not to jeopardize the raw data it could be
implemented in the pipeline software that converts the raw data file
into the final data file for the user in FITS format
(\cite{Pence:2009}), as for S-Cam 3. For the real-time analysis this
can be performed in a parallel route to the operating software, e.g.
using a lookup table to convert pulse height ratio into virtual
pixel to save calculation time. This would provide a highly
desirable real time preliminary image reconstruction.

\section{Conclusion}\label{conclusion}
We have successfully demonstrated operation of an array of DROIDs as
a photon counting and imaging detector in an astronomical
instrument. Although the responsivity of the array was too low for
practical use, the resolving power and imaging capabilities, which
will improve further with increasing responsivity, are already
adequate. From this first system test we have obtained a good
understanding on how to further optimize the system for photon
detection with DROIDs.

\end{document}